\documentstyle[aps,epsfig]{revtex}
 
\begin{document}

\title{\bf Energy dependence of mass distributions in fragmentation}

\author{Emily S.C. Ching$^*$, Y.Y. Yiu and K.F. Lo}
\address{Department of Physics,
The Chinese University of Hong Kong, Shatin, Hong Kong}

\date{\today}
\maketitle

\begin{abstract}

We study fragmentation numerically using a simple model in which
an object is taken to be a set of particles that interact
pairwisely via a Lennard-Jones potential while the effect of the
fragmentation-induced forces is represented by some initial velocities
assigned to the particles. The motion of the particles, which is given
by Newton's laws, is followed by molecular dynamics calculations.
As time evolves, the particles form clusters which are identified as
fragments. The steady-state cumulative distribution of the fragment
masses is studied and found to have an effective power-law region. 
The power-law exponent increases with the energy given to the
particles by the fragmentation-induced forces. This result is 
confirmed by experiments.

\end{abstract}

\vspace{0.5cm}

\hspace{1cm} {keywords: Fragmentation; Mass distributions; Energy dependence}

\vspace{1cm}

\hspace{1cm} $^*$email: ching@phy.cuhk.edu.hk

\newpage

\section{Introduction}
Fragmentation is a common physical process that occurs 
in everyday life and in many areas of science and technology.
When an object experiences an impact or a shock in which
it receives sufficient energy, it will break up into many 
smaller pieces or fragments. Although fragmentation is a 
complex process that involves propagation of many cracks 
and their interaction, 
some simple overall statistical features have been observed.
A large amount of data have been collected which 
indicate that for a wide variety of fragmented objects,
ranging from sand and gravels to asteroids, 
the cumulative distribution of sizes can be represented by a 
power law for some non-negligible range\cite{TK}. 
Such feature was also found in the cumulative distribution
of fragment volumes in the brittle fracture of glass 
spheres\cite{GB61}.

Fragmentation of long glass rods dropped onto the floor\cite{IM}
showed that the distribution of fragment masses changes from lognormal
to one with a power-law region when the dropping 
height is large enough. Oddershede, Dimon and Bohr\cite{ODB} 
studied fragmentation of objects made of different materials 
including gypsum, soap, and potato, and reported that the 
cumulative fragment mass distribution can be fitted by a power law 
with an exponential cutoff. Moreover, the cumulative distribution 
was found to be insensitive to the types of the material with the 
power-law exponent depending only on the morphology of the objects. 
These observations were interpreted\cite{ODB} as evidence that the 
fragmentation process is a self-organized critical phenomenon and 
that the power-law exponent depends only on the effective dimension 
of the objects. In a later study using plates of dry clay,
Meibom and Balslev\cite{MB} observed instead that the cumulative 
fragment mass distribution has two power-law regions and a cutoff, 
with the exponent larger for the smaller fragments. Moreover, the 
two regions were found to be separated by a mass whose value 
increases with the thickness of the clay plates. Based on these 
results, they concluded\cite{MB} that the power-law exponent 
depends on the dimensionality of the original object on the scale 
of the fragment considered and not simply on the global dimensionality 
of the object. Both experimental studies\cite{ODB,MB} indicate that
the power-law exponent increases with dimensionality.

Geometric statistical models\cite{MS,GK} for fragmentation 
have been studied in which the fragments are produced according 
to an assumed distribution of pre-exisiting fracture points, 
lines or planes in one, two or three dimensions. 
The fragment size distribution depends on the form
of the distribution of the pre-existing flaws. For example,
the cumulative fragment size distribution will be 
exponential\cite{GK} for an one-dimensional object if 
the fracture points obey a Poisson distribution.
Another approach is to describe the fragmentation
process by a rate equation\cite{R}. Power-law fragment 
mass distributions can be obtained when the relative breakup 
rate is taken to be a power-law form. A third approach 
considers fragmentation of an object as a partition
of a set of particles. This is mathematically equivalent to the
partition of an integer\cite{SM}. When each partition is assumed 
to be equally likely, the actual distribution would be 
the one that can be realized in the largest number of ways which
can be obtained by maximizing the information entropy\cite{AH,ERJ}.
On the other hand, different weights could be assigned to the
partitions to get different fragment size distributions\cite{M}.

Various models have been introduced to study fragmentation 
numerically\cite{IT,H,KH,NSAS}. Inaoka and Takayasu\cite{IT} modeled
the impact fracture process as a competitive growth of fragments. 
Together with Toyosawa, they found that the fragment mass distribution
depends on how the object is hit when the object has an aspect ratio
other than one. The distribution shows a power law with a flat region
then a cutoff when the object is hit by its thin side but a power law 
with a cutoff when it is hit by its broad side\cite{ITT}. The former
results resemble the experimental observation of two power-law 
regions\cite{MB} discussed eariler.  In the other models, an object is 
described by an assembly of basic constituents or building blocks which 
are connected to each other via elastic springs or beams\cite{H,KH}, or 
by some history-dependent attractive force\cite{NSAS}. The object is 
then subjected to some external force\cite{H,KH} or the constituents be 
given some initial random velocites from a uniform distribution with a 
maximum value\cite{NSAS} and the whole system evolves according to 
Newtonian dynamics. The connection or bond between the constituents is 
taken to be broken when some specific rules are satisfied. Fragment mass 
distributions with a power-law region were found in these numerical 
studies (except when the object is broken under both compression and 
elongation\cite{KH}) and this result is robust to variation in the parameters
of the models.

In this paper, we study how the fragment mass distribution depends on 
the input energy imparted by forces that cause the fragmentation. 
Such a dependence was suggested in Ref.~\cite{NSAS} in which the power-law
exponent was found to change considerably when the maximum velocity in 
the uniform distribution is around 1. However, as discussed in that 
reference, it was not clear that whether the dependence is
continuous as the maximum velocity is varied or whether 
there is a crossover from other values of the maximum velocity 
to the value of 1. To study such dependence in detail, 
we use a simple model of fragmentation in which we can 
vary the input energy easily. The model 
used is described in Sec. II. Results obtained are presented in Sec. III 
and we shall see that there is a continuous change in the power-law 
exponent as the input energy is increased. In Sec. IV, we discuss our 
results and compare them to some experimental observations. 
The paper ends with a conclusion in Sec.~V.

\section{The Model}
A model of fragmentation should contain at least two aspects:
a description of the object to be fractured and a representation 
of the effect of the fracturing forces. The simplest model 
of an object is that it consists of $N$ particles. In order for the
object to remain as a whole piece but without collapsing to a point, 
the particles should attract one another when they are far away 
but repel each other when they are too close. A convenient choice 
of such an interaction is the Lennard-Jones potential: 
\begin{equation}
V(r) = {K_1 \over r^{12}} - {K_2 \over r^{6} }  \ ,
\label{LJ} \end{equation}
with $r$ being the separation between the particles.
To speed up the calculations, the interaction potential $V$
is truncated when $r$ is three times the equilibrium separation
$\sigma$. The parameters $K_1$ and $K_2$ are fixed to be $5$
and $10$ respectively, and $\sigma$ is thus 1~unit.
We have checked that the truncation does not affect the results 
for the fragment mass distributions.
Every particle is taken to have
the same mass of 1 unit. In order to obtain a statistically
significant number of fragments
and at the same time without making $N$ formidably large
computationally, we study two-dimensional objects. The initial
positions of the particles are chosen at random inside a square
whose size is set by the number density~$\rho$. A snapshot of the
initial configuration of the particles is shown in Fig.~1.

A major effect of the impact or shock that causes fragmentation 
is to impart energy to the object.
With the imparted energy, bonds are broken and, as a result, 
the object breaks up into pieces. This effect of the 
fragmentation-induced forces is represented by assigning the 
particles some initial velocities as was done in Ref.\cite{NSAS}. 
In this spirit, different situations of fragmentation can be studied 
by assigning different distributions of initial velocities to the particles. 

The fragmentation process is followed using molecular dynamics 
calculations which were introduced to study fragmentation
by Holian and Grady\cite{HG}. To compare with experiments, we use a 
free boundary condition instead of an expanding periodic boundary
condition as used in their study. Each particle moves according to
Newton's laws of motion. At each time step, the positions and 
velocities are updated for all the particles at the same time by the
Verlet algorithm\cite{GT}. Both the total momentum and the total energy, 
which is the sum of the potential and
kinetic energies, of the particles are conserved during the
fragmentation process. Momentum conservation is automatic in the Verlet
algorithm while energy conservation is enforced in our numerical 
scheme\cite{Yiu}.

\section{Results}
We first report results for the case when the particles
are given some initial radial velocities. The magnitude
of the velocity is proportional to the distance of 
the particle from the center of the square:
\begin{eqnarray}
\nonumber
v_{xi} &=& C x_i \\
v_{yi} &=& C y_i
\label{vel}
\end{eqnarray}
where $x_i$ and $y_i$ are the distances of the $i$-th
particle from the center along the $x$ and $y$-directions.
This velocity distribution was also studied in Ref.\cite{KH,HG},
and might be realized in fragmentation of metal rings that were
forced to undergo rapid radial expansion through 
the application of an impulsive magnetic load\cite{GB} or in
explosive munitions tests\cite{MH}.
Besides the radial velocities, each particle has
also a small thermal random velocity. The thermal kinetic energy
is always less than the kinetic energy of the assigned velocities.

In this model, the input energy imparted by the fracturing forces
can be easily measured by the kinetic energy due to the assigned
initial velocities. Therefore, we vary the value of the
proportionality constant $C$ in (\ref{vel}) to change
the input energy. The initial configuration
of the particles is kept fixed so that the initial potential energy
of the system remains constant. The random thermal energy is also
kept fixed. Thus, a convenient parameter to describe the situations 
of different input energies is the ratio of the initial kinetic 
energy to the absolute value of the initial potential energy of 
the particles, which we denote as $R$.
Snapshots of the configuration of the
particles for $R = 0.43$ at some later times are shown in Fig.~2. 
It can be seen that the particles distribute themselves 
in clusters of various sizes.
Particles that are located within a distance $d= 3.0$~$\sigma$ of
one another are naturally defined to be in one fragment. 
The mass of each fragment is equal to the number of particles in 
the fragment as we have taken the mass of each particle to be one unit.

To study the statistical features of the fragmentation process,
we calculate the cumulative mass distribution.
Denote the
distribution of fragments of mass $m$ by $n(m)$,
the cumulative distribution $F(m)$ is defined as
\begin{equation}
F(m) = \int_m^\infty n(m') dm' 
\label{Fm}
\end{equation}
which measures 
the number of fragments whose mass larger than $m$. 
We calculate $F(m)$ regularly and find that it 
becomes steady after a certain large enough number 
of time steps (about 5000 time steps for a time step of 0.01).
The steady-state cumulative distributions $F(m)$ are analyzed.
All the results reported below refer to the steady-state distributions.

In Fig. 3, we show $F(m)$ for $R = 0.43$ and $N = 4200$. 
A short effective power-law region which extends for
about a decade can be observed:
\begin{equation}
F(m) \sim m^{-\alpha}
\label{powerlaw}
\end{equation}
for intermediate fragment masses. The power-law exponent $\alpha$ is found
to be $0.41$ by fitting the best straight line region in the log-log plot.
This power-law region is verified by the existence of a plateau in the 
plot of $F(m)m^{\alpha}$ versus $m$ as shown in the inset of Fig.~3.
We have checked that the results do not
change much if $d$ is in the range of $1.5$ to $3.0$~$\sigma$.
The extent of the power-law region is longer when we use a larger $N$.
It follows from (\ref{Fm}) that $n(m)$ 
also has a power-law region whose exponent 
is given by $\alpha + 1$, i.e., $n(m) \sim m^{-(\alpha +1)}$.

In Fig. 4, we plot the cumulative mass distributions for various
values of $R$ from 0.43 to 6.49. As $R$ increases, $F(m)$ changes. 
The fragment mass distribution is thus not universal. Specifically, 
the power-law region shrinks and the exponent $\alpha$ increases as $R$ 
increases. (For the largest value $R=6.49$, the power-law region
barely exists.) The dependence of $\alpha$ on $R$ is plotted in Fig.~5.
As the input energy increases, we find the average fragment mass 
decreases as expected. In Ref.\cite{HG}, $F(m)$ is fitted by a sum of two
exponentials. We find that this form works for small values of $R$ but
not for intermediate values of $R$.
A decrease in the average fragment mass with $C$ was
also reported in Ref.~\cite{HG} but the dependence of $F(m)$ 
on $C$ was not presented.

We have tested the robustness of our result by changing various
parameters in the model~\cite{Yiu}. Initial configurations 
corresponding to different values of the number density $\rho$ 
have been studied. Different values of $N$ have been used. 
Furthermore, particles initially confined inside a circular 
region have also been studied. The same result that the exponent 
of the power-law region of $F(m)$ increases with $R$ (for values of 
$R$ not too large; see below) is observed in all these different cases.

Seeing the exponent increases with the input 
energy, a natural question is whether the power-law behavior
will continue in the limit of very large input energy. 
In our numerical study, we find that when $R$ is larger than 
a certain value, there is no longer
any discernable power-law region in $F(m)$. 
In the limit of very large $R$ ($R > 100$), 
$F(m)$ becomes exponential. However, for such large values 
of $R$, the average mass of the fragments is only about 2. This suggests 
that a larger $N$ has to be used for such explorations. 
Since calculations with a large $N$ are computationally expensive,
this question would be more easily addressed by experimental 
studies\cite{CCX}.

We have also investigated the effect of the form of the 
initial assigned velocities on the mass distribution by studying
the case when only velocities with random direction 
are assigned to the particles\cite{NSAS}. This case corresponds 
physically to the object being heated up. When the velocities are 
small or the temperature is low, the object remains mainly as a 
whole piece as expected. As the magnitude of the velocities increases,
we find that an effective power-law region also develops 
in $F(m)$ but with a shorter extent as compared to that 
for radial initial velocities. It would be 
interesting to study experimentally whether power law 
could indeed be observed in the distributions of fragments 
resulting from thermal dissociation. 

\section{Discussion}
A possible dependence of the power-law exponent  
on input energy has been hinted by earlier experiments\cite{IM,K}.
To test this result directly, we have performed experiments 
of long glass rods dropped onto the floor similar to those 
reported in Ref.\cite{IM} specifically to check the dependence
of the fragment distributions on the input energy. 
Details of the experiment will be reported elsewhere\cite{CCX}.
Here, we shall outline the main experimental observations and 
compare them with our simulation results.  

Potential energy is lost as the glass rod is dropped from a height
onto the floor. If all the energy is imparted to
the glass rod when it reaches the floor and causes it to break up
into fragments, then the input energy will be proportional to the
height of the fall. Thus, the glass rods are dropped 
from various heights and the corresponding distributions 
of the fragment masses analyzed. The power-law region is again
found to shrink as the height increases. Moreover, the exponent 
$\alpha$ increases from about $0.3$~to~$0.8$ for heights of fall 
ranging between 1.2~m to 10.6~m similar to that observed in the simulation. 

If the power law in the fragment mass distribution is a result of some 
self-organizing effect of the dynamics of the fragmentation process, 
one would expect that the exponent should not depend on how much the 
system is driven or, in other words, how much the input energy is. 
Thus, a dependence of the exponent on the input energy suggests 
that the fragmentation of glass rods on dropping onto the floor 
is not a self-organized critical phenomenon, contrary to that 
suggested in Ref.\cite{ODB}.

A dependence of the power-law on input energy
was also found in an analytical model of fragmentation 
recently studied by Marsili and Zhang\cite{MZ}. In their model, 
fragmentation is assumed to happen in a hierarchical order and 
at each level of the hierarchy, a fragment will break further or
not into two smaller pieces depending on whether or not its energy 
density exceeds a certain threshold. They found that the distribution
of fragment sizes has an effective power law with the effective 
exponent increases with the input energy, and equals to the 
dimension of the object in the limit of very large input energy. 
There is, however, no obvious hierarchical order 
in the breaking of the object in our model. 

\section{Conclusion}
We have studied a simple model of fragmentation numerically
using molecular dynamics calculations.
The two essential physical ingredients of the model are: (i)
the consideration of an object as a set of particles that interact 
with each other pairwisely via a truncated Lennard-Jones potential 
and (ii) the approximation of the effect of abruptly applied 
fracturing forces by some specified initial velocities assigned 
to the particles. Although the model is simple,
the fragment mass distribution obtained displays some features
observed in experiments. First, there is an effective power-law 
region for a range of input energies. The origin of such 
power laws can be seen as a result 
of the competition between the forces that tend to break up 
an object and the cohesive forces that keep the object as a whole
when the two kinds of forces do not differ too much in magnitude. 
Second, the effective exponent increases with the input energy, 
as observed in the experiments of dropping glass rods onto 
the floor\cite{CCX}. This second feature suggests
that the impact fragmentation process is not a self-organizing
phenomenon. Our study thus sheds light on why power-law behaviour is 
so overwhelmingly persistent in the observation of fragment distributions 
and why a range of power-law exponents has been reported. 

\acknowledgements
We thank H. J. Herrmann, N. H. Kwong, C. X. Wang, and K. Q. Xia  
for discussions. E. S. C. C.  is grateful to A. Libchaber for his
encouragement. This work is supported by  
a Direct Grant of the Chinese University of Hong Kong, and also
in part by the Hong Kong Research Grant Council (Grant no. 315/96P).

\newpage

\centerline{FIGURE CAPTIONS}

\vspace{12pt}

\noindent
FIG. 1. The initial configuration of the particles for $N=4200$
and $\rho = 0.61$. The coordinates are measured in arbitrary units;
for comparison, the equilibrium separation between two particles is
1 unit.

\vspace{12pt}

\noindent
FIG. 2. 
The configuration of the particles at (a) 1000 and (b) 5000 
time steps for the same parameters in Fig.~1 and $R = 0.43$.

\vspace{12pt} 
 
\noindent 
FIG. 3. The cumulative distribution of fragment masses $F(m)$ versus
$m$ for the same parameters in Fig.~2. $m$ is measured in terms
of the mass of 1 particle which is taken to be 1 unit.
An effective power-law region of about a decade is observed 
and is fitted by the solid line with an exponent $\alpha$ 
equals to $0.41$. In the inset, we plot $F(m)m^\alpha$ versus
$m$, the existence of a plateau region supports the power-law 
behavior.

\vspace{12pt} 
 
\noindent 
FIG. 4. The cumulative distribution of fragment masses $F(m)$ versus
$m$ for various values of $R$. $R = 0.43$~(circles),
$R=1.19$~(squares), $R=2.45$~(rhombuses),
$R=4.22$~(triangles), and $R=6.49$~(triangles). 
The power-law region shrinks and the exponent $\alpha$ 
increases generally as $R$ increases.

\vspace{12pt} 
 
\noindent 
FIG. 5. The $R$-dependence of the power-law exponent $\alpha$
for the cases shown in Fig.~4.

\newpage

\begin{figure}
\epsfig{file=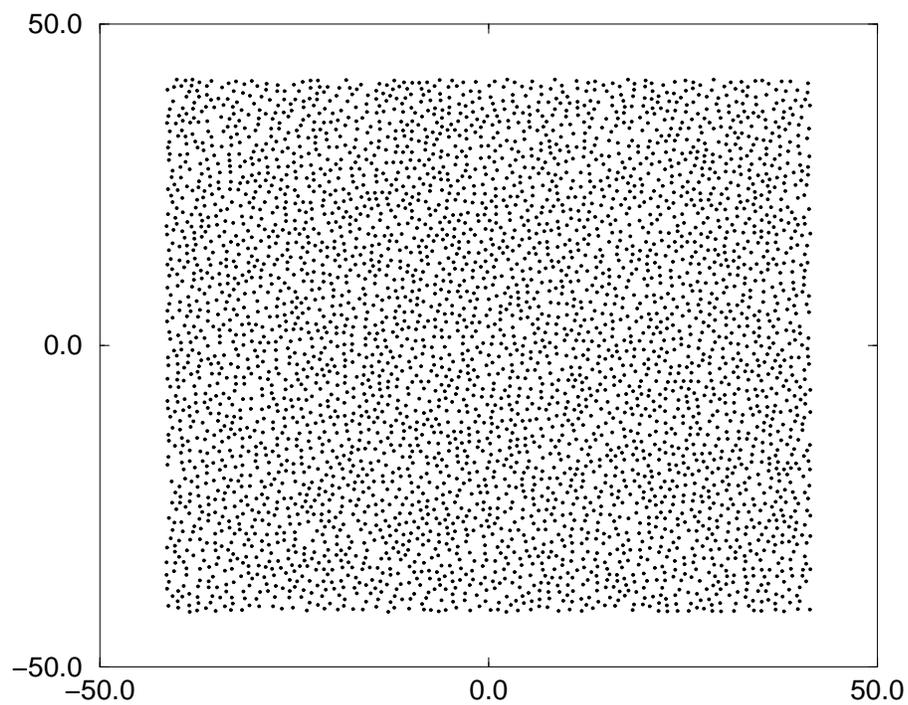,width=12cm}
\caption{The initial configuration of the particles
for $N=4200$ and $\rho=0.61$.
The coordinates are measured in arbitrary units;
for comparison, the equilibrium separation between two particles is
1 unit.}
\end{figure}
\begin{figure}
\epsfig{file=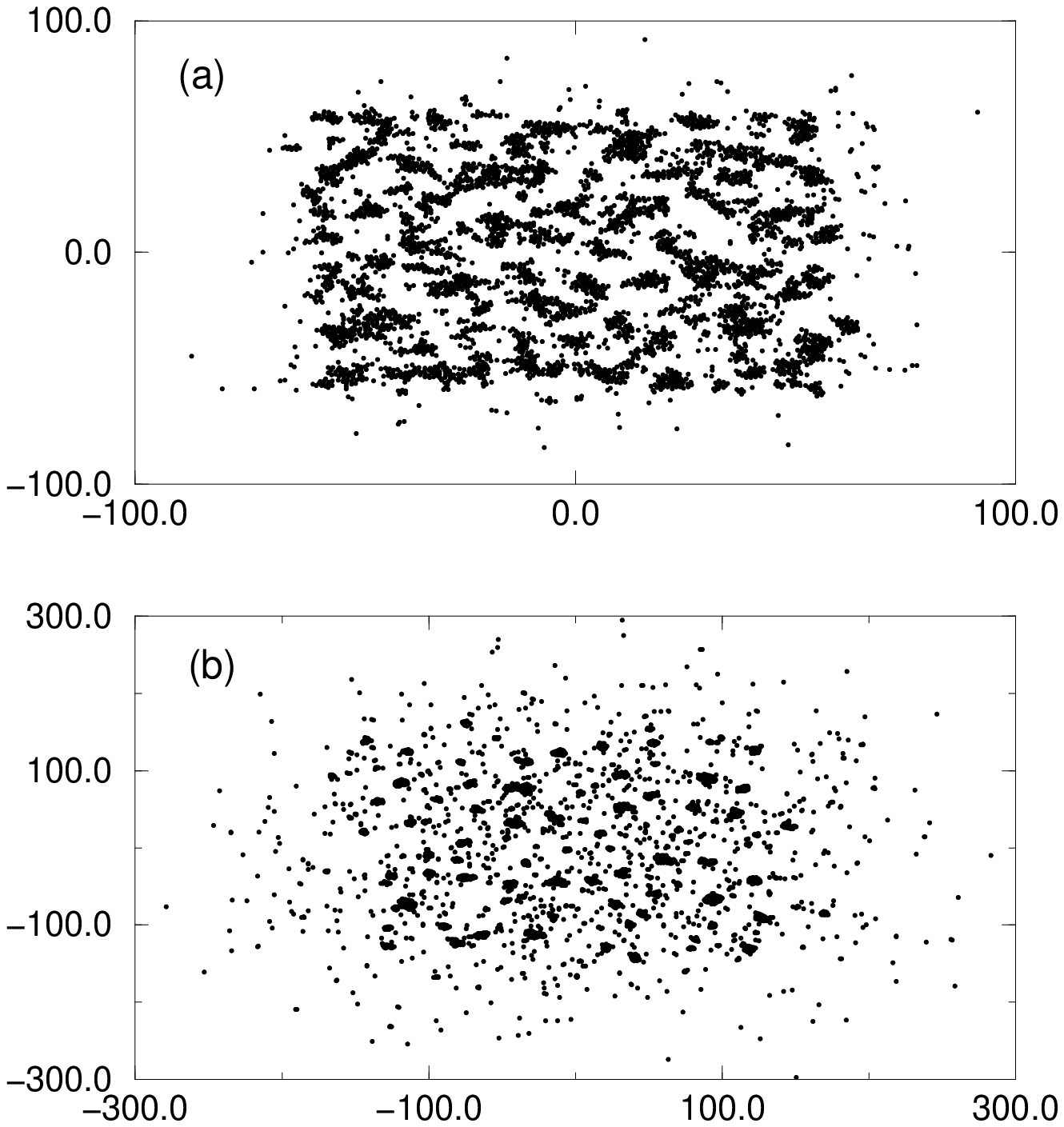,height=19cm,width=13cm}
\caption{The configuration of the particles at (a) 1000 and (b) 5000 
time steps for the same parameters in Fig.~1 and $R = 0.43$.}
\end{figure}
\begin{figure}
\epsfig{file=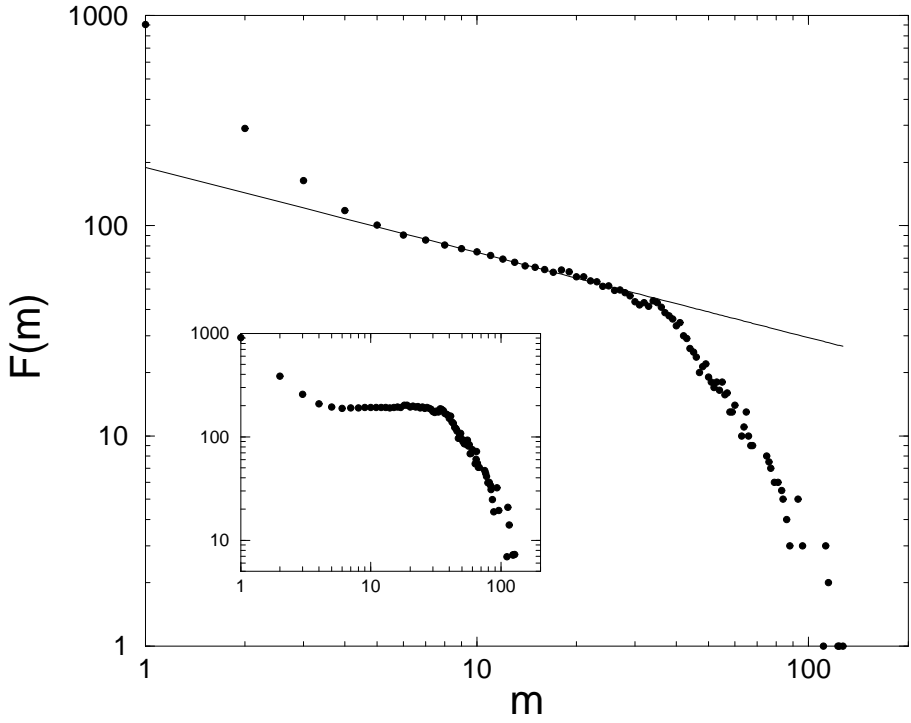,width=12cm}
\caption{The cumulative distribution of fragment masses $F(m)$ versus
$m$ for the same parameters in Fig.~2. $m$ is measured in terms
of the mass of 1 particle which is taken to be 1 unit.
An effective power-law region of about a decade is observed 
and is fitted by the solid line with an exponent $\alpha$ 
equals to $0.41$. In the inset, we plot $F(m)m^\alpha$ versus
$m$, the existence of a plateau region supports the power-law 
behavior.}
\end{figure}
\begin{figure}
\epsfig{file=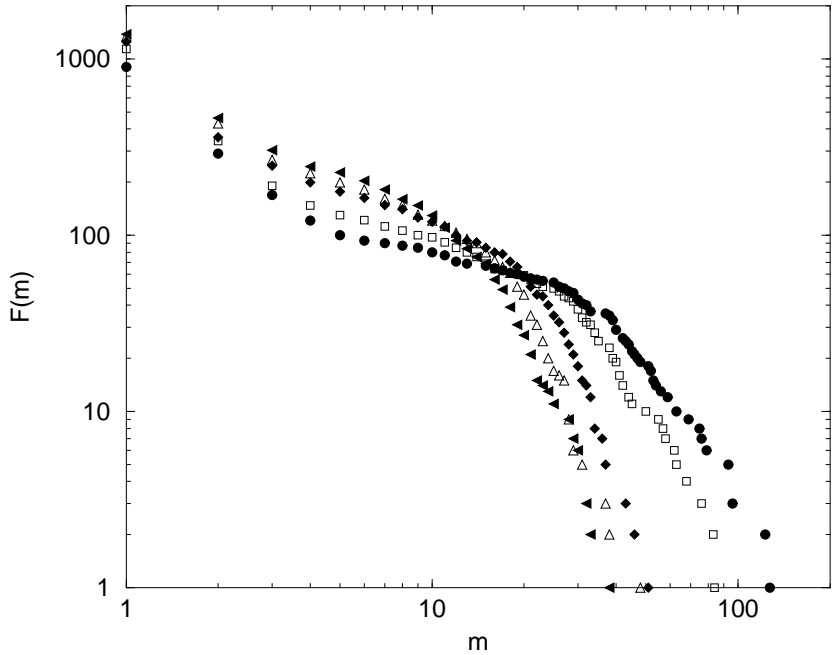,width=11cm}
\caption{
The cumulative distribution of fragment masses $F(m)$ versus
$m$ for various values of $R$. $R = 0.43$~(circles),
$R=1.19$~(squares), $R=2.45$~(rhombuses),
$R=4.22$~(triangles), and $R=6.49$~(triangles). 
The power-law region shrinks and the exponent $\alpha$ 
increases generally as $R$ increases.}
\end{figure}
\begin{figure}
\epsfig{file=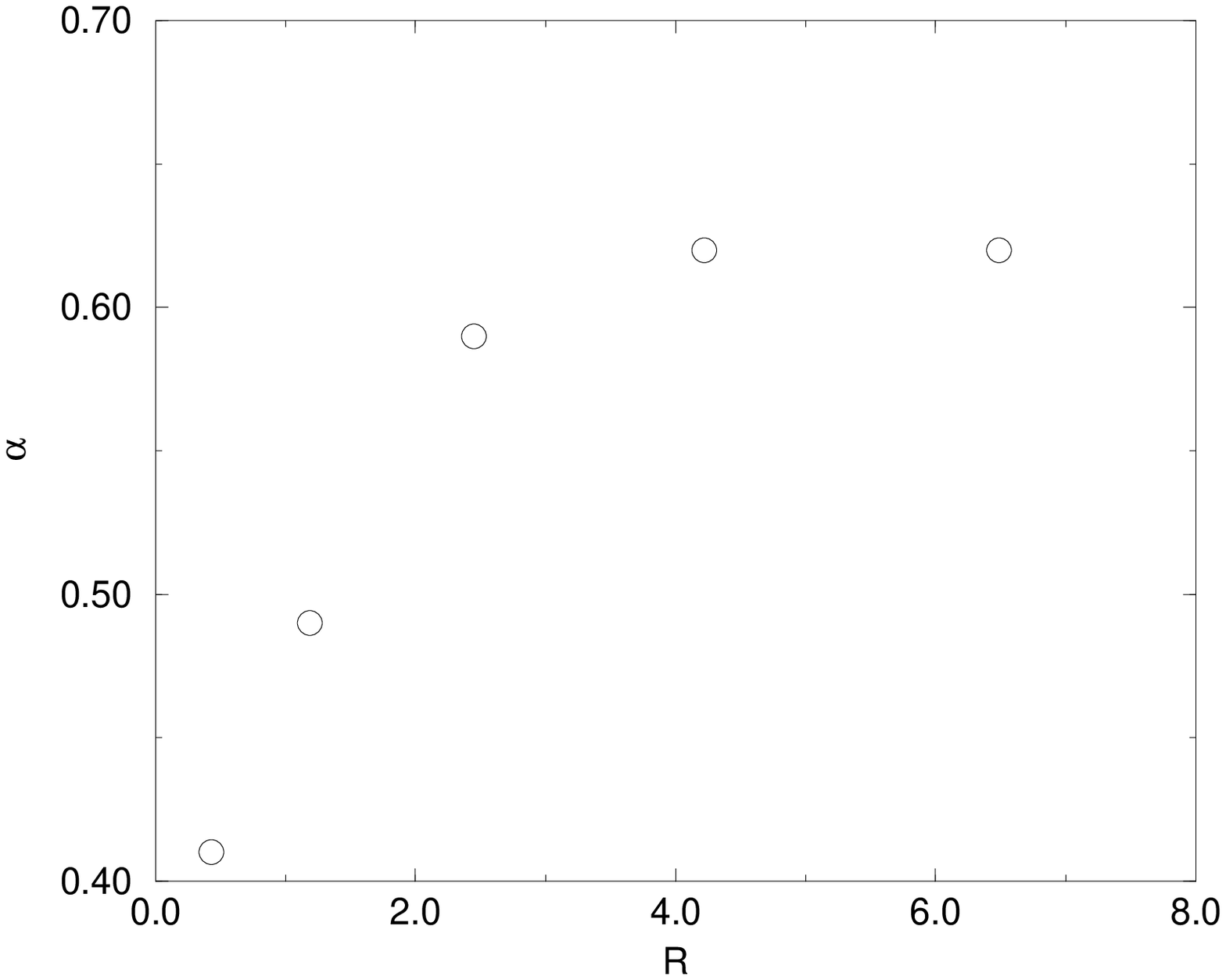,width=11cm}
\caption{
The $R$-dependence of the power-law exponent $\alpha$
for the cases shown in Fig.~4.}
\end{figure}

\begin{references}

\bibitem{TK} D.L. Turcotte, J. Geophys. Res. B 91 (1986) 1921;
G. Korvin, Fractal Models in the Earth Sciences
(Elsevier Science, 1992).

\bibitem{GB61} J.J. Gilvarry and B.H. Bergstrom, J. Appl. Phys. 
32 (1961) 400.

\bibitem{IM} T. Ishii and M. Matsushita, J. Phys. Soc. Jpn. 61
(1992) 3474.

\bibitem{ODB} L. Oddershede, P. Dimon, and J. Bohr, Phys. Rev. Lett.
71 (1993) 3107.

\bibitem{MB} A. Meibom and I. Balslev, Phys. Rev. Lett. 76
(1996) 2492.

\bibitem{MS} M. Matsushita and K. Sumida, Chuo Univ, 31
(1981) 69.

\bibitem{GK} D.E. Grady and M.E. Kipp, J. Appl. Phys. 58
(1985) 1221.

\bibitem{R} H.J. Herrmann and S. Roux, eds., 
Statistical Models for the Fracture of Disordered Media 
(Elsevier Science, New York, 1990).

\bibitem{SM} L.G. Sobotka and L.G. Moretto, Phys. Rev. C 31 
(1985) 668.

\bibitem{AH} J. Aichelin and J. Huefner, Phys. Lett. 136 B
(1984) 15.

\bibitem{ERJ} R. Englman, N. Rivier, and Z. Jaeger, Phil. Mag. B 56 
(1987) 751.

\bibitem{M} A.Z. Mekjian, Phys. Rev. Lett. 64 (1990) 2125.

\bibitem{IT} H. Inaoka and H. Takayasu, Physica A 229 (1996) 5.

\bibitem{H} Y. Hayakawa, Phys. Rev. B 53 (1996) 14828.

\bibitem{KH} F. Kun and H.J. Herrmann, 
Comp. Meth. Appl. Mech. Eng. 138 (1996) 3.

\bibitem{NSAS} U. Naftaly, M. Schwartz, A. Aharony, 
and D. Stauffer, J. Phys. A 24 (1991) L1175.

\bibitem{ITT} H. Inaoka, E. Toyosawa, and H. Takayasu, Phys. Rev. Lett.
78 (1997) 3455.

\bibitem{HG} B.L. Holian and D.E. Grady, Phys. Rev. Lett. 60
(1988) 1355.

\bibitem{GT} See, e.g., H. Gould and J. Tobochnik, An Introduction to
Computer Simulation Methods (Addison-Wesley, 1988).

\bibitem{Yiu} Y.Y. Yiu, M.Phil. Thesis, The Chinese University of 
Hong Kong, unpublished (1997).

\bibitem{GB} D.E. Grady and D.A. Benson, Exp. Mech. 23
(1983) 393.

\bibitem{MH} W. Mock, Jr. and  W.H. Holt, J. Appl. Phys. 54
(1983) 2344.

\bibitem{K} T. Kadono, Phys. Rev. Lett. 78 (1997) 1444.

\bibitem{CCX} E.S.C. Ching, M.P. Chan and K.Q. Xia,
The Chinese University of Hong Kong preprint.

\bibitem{MZ} M. Marsili and Y.-C. Zhang, Phys. Rev. Lett. 77 
(1996) 3577.
\end{references}
\end{document}